\PassOptionsToPackage{table}{xcolor}
\PassOptionsToPackage{prologue,dvipsnames}{xcolor}

\documentclass[sigconf,authorversion,nonacm]{acmart}

\usepackage{booktabs,siunitx}
\usepackage{array}
\usepackage{enumerate}
\usepackage{enumitem}
\setlist{leftmargin=5mm}
\usepackage[framemethod=TikZ]{mdframed}
\mdfsetup{skipabove=2pt}

\newcommand{\rqone}[1]{Can experience-aware oversampling models correctly generate comments?}
\newcommand{\rqtwo}[1]{Can experience-aware oversampling models generate more informative comments?}
\newcommand{\rqthree}[1]{What kind of comments do experience-aware oversampling models generate?}

\AtBeginDocument{%
  \providecommand\BibTeX{{%
    \normalfont B\kern-0.5em{\scshape i\kern-0.25em b}\kern-0.8em\TeX}}}

\ccsdesc[500]{Software creation and management}
\ccsdesc[500]{Machine translation}

\keywords{Code Review, Review Comments, Neural Machine Translation}

\copyrightyear{2024}
\acmYear{2024}
\setcopyright{rightsretained} 

\acmConference[MSR '24]{21st International Conference on Mining Software Repositories}{April 15--16, 2024}{Lisbon, Portugal}

\acmBooktitle{21st International Conference on Mining Software Repositories (MSR '24), April 15--16, 2024, Lisbon, Portugal}

\acmDOI{10.1145/3643991.3644910}

\acmISBN{979-8-4007-0587-8/24/04}

\begin{document}
\title{Improving Automated Code Reviews: Learning from Experience}

\author{Hong Yi Lin}
\affiliation{%
  \institution{The University of Melbourne}
  \city{Melbourne}
  \country{Australia}
}
\email{holin2@student.unimelb.edu.au}

\author{Patanamon Thongtanunam}
\affiliation{%
  \institution{The University of Melbourne}
  \city{Melbourne}
  \country{Australia}
}
\email{patanamon.t@unimelb.edu.au }

\author{Christoph Treude}
\affiliation{%
  \institution{Singapore Management University}
  \city{Singapore}
  \country{Singapore}
}
\email{ctreude@smu.edu.sg}

\author{Wachiraphan Charoenwet}
\affiliation{%
  \institution{The University of Melbourne}
  \city{Melbourne}
  \country{Australia}
}
\email{wcharoenwet@student.unimelb.edu.au}

\begin{abstract}
Modern code review is a critical quality assurance process that is widely adopted in both industry and open source software environments.
This process can help newcomers learn from the feedback of experienced reviewers; however, it often brings a large workload and stress to reviewers.
To alleviate this burden, the field of automated code reviews aims to automate the process, teaching large language models to provide reviews on submitted code, just as a human would.
A recent approach pre-trained and fine-tuned the code intelligent language model on a large-scale code review corpus.
However, such techniques did not fully utilise quality reviews amongst the training data. 
Indeed, reviewers with a higher level of experience or familiarity with the code will likely provide deeper insights than the others.
In this study, we set out to investigate whether higher-quality reviews can be generated from automated code review models that are trained based on an experience-aware oversampling technique.
Through our quantitative and qualitative evaluation, we find that experience-aware oversampling can increase the correctness, level of information, and meaningfulness of reviews generated by the current state-of-the-art model without introducing new data.
The results suggest that a vast amount of high-quality reviews are underutilised with current training strategies.
This work sheds light on resource-efficient ways to boost automated code review models.
\end{abstract}

\maketitle

\section{Introduction}
Modern code review is a widely adopted quality assurance process that leverages the expertise of experienced reviewers to help maintain the source code.
This activity provides an avenue in which novice and new team members can learn from the feedback of experienced reviewers and improve the overall quality of code changes~\cite{howdevelopers,characteristics_microsoft,samsung,turzo2023makes}.
In the absence of code reviews, code changes become more defect-prone~\cite{mcintosh2015,Kononenko2015,thongtanunam2016ownership}.

In practice, code reviews are demanding workloads for reviewers as they require large amounts of time and attention.
To help alleviate this stressful and time-consuming process, several works attempt to automate the practice by leveraging large language models to imitate reviewers~\cite{tufano2019learning,tufano2021towards,tufano2022using,thongtanunam2022autotransform,li2022auger,li2022codeeditor,dact,structure}.
Most recently, Li et al.~\cite{li2022automating} proposed CodeReviewer, a pre-trained code model on the largest code review dataset that achieved state-of-the-art performance.
However, such methods still treat all review examples (i.e., training data) as equal in quality, irrespective of the experience of the reviewer behind the comment. 

In this work, we hypothesise that spending more training on experienced reviewers' examples can help the model pay more attention to critical issues within code changes and communicate better insights into underlying problems, resulting in better quality reviews.
Rather than acquiring more data to train the model, we treat the experienced reviewers' examples as low-resource data~\cite{currey-etal-2020-distilling, przystupa-abdul-mageed-2019-neural,tan-etal-2019-multilingual}.
Thus, we use oversampling to over-represent target examples during training, such that these examples yield more influence over the model's behaviour, enabling higher-quality review generation.

To investigate this, we fine-tune CodeReviewer~\cite{li2022automating} with oversampled experienced reviewers' examples to automatically generate code review comments.
Then we evaluate our experience-aware oversampling models in terms of correctness~\cite{li2022automating, tufano2022using}, level of information~\cite{li2022automating}, and meaningfulness~\cite{characteristics_microsoft, microsoftbugs, samsung, howdevelopers}.
% Leveraging CodeReviewer \cite{li2022automating} and their benchmark dataset, we demonstrate the ability to generate higher-quality reviews in terms of correctness, level of information and meaningfulness, without introducing new data.
% More specifically, we demonstrate that a higher quality of reviews can be generated by oversampling examples from reviewers who were experienced in either authoring, reviewing or both, allowing extra influence over model training.
Through our quantitative and qualitative evaluation, we found that our experience-aware oversampling models can generate more comments that are semantically correct (16\%-21\%), and more applicable suggestions to the proposed changes (32\%-34\%) with explanation (9\%-16\%) than the original model, which achieves 15\% for semantic correctness, 22\% for applicable suggestions, and 4\% for explanation.
% 16\%-23\% of the comments generated by our experience-aware oversampling models were semantically correct,
% whilst 14\% of the comments from the original model of CodeReviewer were semantically correct. 
% 31\%-33\% of the comments by our experience-aware oversampling models provided applicable suggestions to the proposed code changes, while the original model achieved only 20\%.
% Additionally, our models were able to provide explanations within the applicable comments 9-15\% of the time, as opposed to 4\% by the original model.
Lastly, we discovered that our models could provide comments for various issues, especially critical ones related to logic, validation, and resources.
These results suggest that a higher quality of reviews can be generated by oversampling examples from experienced reviewers, boosting the automated code review performance without introducing new data.

\section{Related work and Motivation}

\textbf{Automated code reviews.} 
In its earliest form, the proposed task was to directly refine the submitted pre-review code changes to an improved post-review version, i.e., ($M_{pre}$$\rightarrow$$M_{post}$) method pairs mined from the Gerrit code review tool~\cite{tufano2019learning}. 
% More specifically, the Recurrent Neural Network was utilised to facilitate the transformation between ($M_{pre}$$\rightarrow$$M_{post}$) method pairs mined from a code review tool (i.e., Gerrit).
Later, Tufano et al.~\cite{tufano2021towards} tested the vanilla transformer~\cite{vaswani2017attention} on a multimodal input scenario ($M_{pre}$,$R_{nl}$$\rightarrow$$M_{post}$), where $R_{nl}$ represents a natural language comment that helps guide the code refinement.
% With the advent of transfer learning, studies have also found software artifacts focused pre-training \cite{tufano2022using,li2022codeeditor} to be effective.
\begin{comment}
Mirroring the advancements from natural language processing, Tufano et al. \cite{tufano2022using} found large improvements by pre-training the Text-to-Text Transfer Transformer \cite{raffel2020exploring} on the Stack Overflow dump \cite{stackoverflow} and CodeSearchNet \cite{husain2019codesearchnet} before performing code refinement as a downstream task.
On a similar front, Li et al. \cite{li2022codeeditor} were able to boost performances by pre-training CodeT5 \cite{wang2021codet5} on synthetic code improvement data generated by CodeGPT \cite{lu2021codexglue}. 
\end{comment}
Other studies also investigated different techniques to improve performance. 
Thongtanunam et al.~\cite{thongtanunam2022autotransform} used subword tokenisation to unlock the ability to handle previously unseen tokens appearing in $M_{post}$, whilst others found benefits in using code diff~\cite{dact} and structure information~\cite{structure}.

\textbf{Review comment generation.} 
As the ($M_{pre}$,$R_{nl}$$\rightarrow$$M_{post}$) form of code refinement still relies on the comment $R_{nl}$ of a human reviewer, recent techniques~\cite{li2022auger,li2022automating} have focused on incorporating the ability to perform ($M_{pre}$$\rightarrow$$R_{nl}$) review comment generation~\cite{tufano2022using}.
Review comment generation embodies the core task of automated code reviews, where the model needs to identify the exact issue within the submitted code $M_{pre}$ from an expansive potential problem space and output a useful and detailed comment $R_{nl}$ that will assist a human developer in improving the quality of code changes for $M_{post}$. 
Although efforts have been made to filter out noisy and unrelated comments~\cite{tufano2021towards,tufano2022using, li2022auger, li2022automating}, previous works have not focused on the variation in review quality within the datasets.

\textbf{Reviewer experience and comment quality.} 
At the core of code review comment quality, having a wealth of experience enables reviewers to provide better insights. Mozilla core developers~\cite{howdevelopers} argue that a meaningful code review needs to provide more valuable feedback than mere suggestions on code formatting and style.
Moreover, they also argue that feedback from experienced reviewers is preferred as these experienced reviewers can leverage their understanding of the codebase to share insights on what could break and what can be re-used.

Microsoft developers~\cite{characteristics_microsoft} stated that useful comments identify functional and validation issues.
% , they facilitate knowledge sharing regarding project design, constraints and other available tools that could be utilised.
% Issues regarding maintainability, such as style and readability were often only considered somewhat useful.
Developers at Samsung Research Bangladesh~\cite{samsung} suggested that comments related to optimisation, redundant code, corner cases, code integration, deprecated features, and coding standards were also useful.
In a study with OpenDev developers~\cite{turzo2023makes}, defects, code improvement opportunities, and alternative solutions were considered the primary usefulness criteria.
% whilst code maintainability, knowledge sharing and helping relationship formations were considered as secondary criteria.
% In a study with OpenDev developers \cite{turzo2023makes}, defects, code improvement opportunities and alternative solutions were considered as a primary usefulness criteria, whilst code maintainability, knowledge sharing and helping relationship formations were considered as a secondary criteria.
% Functional defects, validation and logical were ranked as the top three most useful types of comments.
% Supporting the findings of past work, they also discovered a positive relationship between reviewer experience and the usefulness of the code review.
Furthermore,~\cite{characteristics_microsoft} report that reviewers who had authoring experience with the file under review had a greater number of useful comments.
As studies in both industry~\cite{characteristics_microsoft, microsoftbugs, revhelper, samsung} and open-source environments~\cite{Kononenko2015, howdevelopers, turzo2023makes} converge on this finding, it becomes evident that the experienced reviewer demographic requires additional attention.

\begin{figure*}[ht] 
    \centering
    \includegraphics[width=\textwidth, clip]{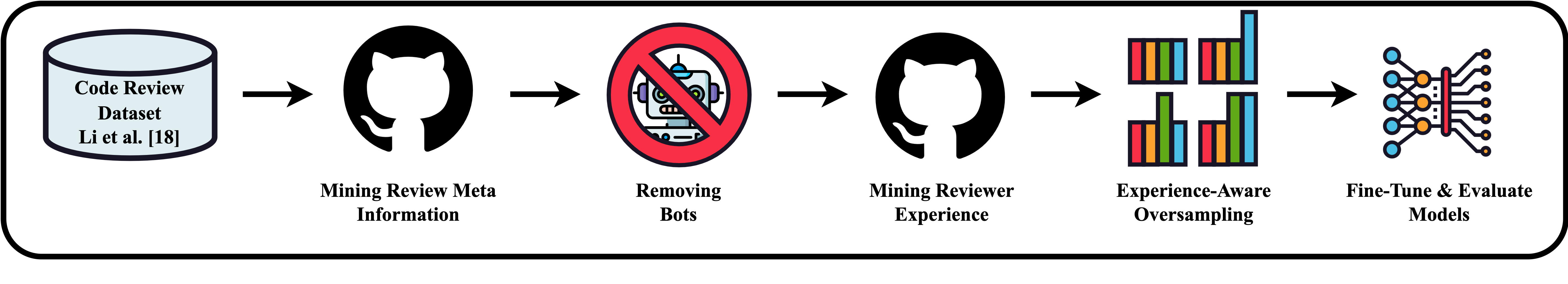} 
    \caption{The Process of Creating Experience-Aware Automated Code Review Models}
    \label{fig:process}
\end{figure*}

\section{Study Design}

\subsection{Research Questions}
Following the notion that review quality is associated with reviewer experience~\cite{characteristics_microsoft, microsoftbugs, revhelper, samsung,Kononenko2015, howdevelopers, turzo2023makes}, we argue that spending more training on experienced reviewers' examples can help the model improve the quality of reviews.
To do so, we treat the experienced reviewers' examples as a low-resource corpus~\cite{currey-etal-2020-distilling, przystupa-abdul-mageed-2019-neural,tan-etal-2019-multilingual}.
% Following the notion that review quality is associated with reviewer experience, we look to investigate if this theory could be embedded into the training of automated code review models.
As such, we propose an experience-aware oversampling approach, where we train a model by targeting reviewers with high authoring and/or reviewing experience and overrepresent their examples during training.
With this approach, the oversampled examples would yield more influence over the model's behaviour, enabling higher-quality review generation.
Since review comment generation embodies the core task of automated code reviews, in this work, we focus on the effectiveness on \textit{code review comment generation}.  
To evaluate our approach, we formulate the following research questions.
\\
\begin{mdframed}[linewidth=0.25mm,roundcorner=12pt, backgroundcolor=black!10]
\textit{\textbf{RQ1:} \rqone } \\
\textit{\textbf{RQ2:} \rqtwo} \\
\textit{\textbf{RQ3:} \rqthree }
\end{mdframed}

RQ1 evaluates the correctness of the generated comment against the ground truth~\cite{li2022automating, tufano2022using}.
RQ2 evaluates the models' ability to generate actionable and understandable reviews~\cite{li2022automating}.
RQ3 evaluates the models' ability to generate review comments that target more critical issues as expected by developers~\cite{characteristics_microsoft, microsoftbugs, samsung, howdevelopers}.

\subsection{Reviewer Experience Heuristics}
To identify experienced reviewers' examples, we calculate ownership metrics that measure the experience (i.e., familiarity with the codebase) of individual reviewers.
We leverage traditional ownership metrics from both the authoring~\cite{bird2011don} and reviewing~\cite{thongtanunam2016ownership} perspective to represent the reviewer's experience.
We calculate the code ownership of a reviewer at the repository level to fit the review environment of the studied data (i.e., GitHub).
% Different to past research, which calculated ownership metrics on a module level, our metrics are adapted to the repository level to fit the GitHub environment.
This level of granularity allows us to capture overall reviewer experience in the project, as target file experience~\cite{characteristics_microsoft,revhelper, samsung} can often be inaccurate due to deletion.
The authoring based \textbf{Authoring Code Ownership (ACO)} metric is calculated as $ ACO(D,R) = \frac{\alpha(D,R)}{C(R)}$
where $\alpha(D,R)$ is the number of commits the reviewer $D$ has contributed to the repository $R$ and $C(R)$ is the total number of commits to the repository.
The \textbf{Review-Specific Ownership (RSO)} metric is calculated as $ RSO(D,R) = \frac{r(D,R)}{\rho(R)}$
where $r(D,R)$ is the number of closed pull request reviews in a repository $R$ for which the reviewer $D$ has provided comments and $\rho(R)$ is the total number of closed pull request reviews that have been conducted in the repository $R$.

% Different to past work \cite{thongtanunam2016ownership}, we do not normalise across the number of reviewers that have participated in each review as we are not exhaustively considering all reviewers within the repositories.
Following the traditional approaches, reviewers with ACO $\ge$ 5\% are considered \textbf{\textit{major authors}}, whilst those with ACO $<$ 5\% are considered \textbf{\textit{minor authors}}~\cite{bird2011don}.
Similarly, those with RSO $\ge$ 5\% are considered \textbf{\textit{major reviewers}}, whilst those with RSO $<$ 5\% are considered \textbf{\textit{minor reviewers}}~\cite{thongtanunam2016ownership}.

\subsection{Data Preparation}
% This section details the construction of the code review dataset.

\textbf{Dataset selection.} 
We used the datasets of Li et al.~\cite{li2022automating} who introduced the largest multilingual code review dataset to date.
This dataset was mined from GitHub pull requests and contains code reviews from the top 10k most starred projects.
% for of each of the nine most popular programming languages. 
The training set was built from repositories with more than 2,500 PRs, while the validation and test sets were built from those with [1,500,2,500) PRs.
Li et al.~\cite{li2022automating} provide three separate datasets for three standalone automation tasks, i.e., code change quality estimation ($M_{pre} \rightarrow revise?$), review comment generation ($M_{pre} \rightarrow R_{nl}$), and code refinement ($M_{pre},R_{nl} \rightarrow M_{post}$). 
Since we need to further mine the ownership of reviewers from GitHub repositories, we used the code refinement dataset which still retained PR IDs along with repository names, and we repurposed this dataset for the task of review comment generation, i.e., ($M_{pre} \rightarrow R_{nl}$). 
The original sizes for the training, validation, and test sets were 150,406 comments, 13,103 comments, and 13,104 comments, respectively.

\textbf{Mining review meta information.} 
To gather the meta information of the reviews for ownership calculation, we used the GitHub REST API through PyGithub\footnote{\url{https://github.com/PyGithub/PyGithub}} to retrieve the original pull requests. 
Since there can be many review comments on a single pull request, we obtained only the code review comment that exactly matched the comments in the dataset.
% scanned through each review comment and matched them to the natural language comment found in the dataset.
% We lose 683 examples due to deletion.
For each of the code review comments obtained, we extracted the username of the reviewer and the time of the comment.
In total, we identified 10,583 reviewer accounts in the training set and 2,763 reviewer accounts in the validation and test set, covering the period between 2011 and 2022.

\textbf{Removing bots.} 
While reviews from bots (e.g., CI bots, style checkers) do not harm the model, the main goal of automated code reviews is to replicate human reviews as a means to complement traditional tools.
We utilise robust heuristics~\cite{EnsBoD} to 1) remove accounts with the \textit{"bot"} suffix in their username~\cite{powerofbots} and 2) remove accounts within an established list of bots~\cite{golzadeh2021ground}. 
In total, we removed 1,207 comments from nine bots in the training set and 96 comments from five bots in the validation and test set.
Similar to previous work~\cite{li2022automating}, we removed comments that only suggest code without providing any natural language comment.
\begin{comment}
We found 7322, 632 and 639 reviews in the training, validation and test sets that reflect this type of review.
This subset of data caused the model to overfit and exhibit mode collapse as they contain large portions of the submitted code. 
We observed that the model learns to only copy and paste segments of the input code verbatim, losing its natural language generation capability. 
\end{comment}
The final training, validation, and test sets have sizes of 141,259 comments, 12,406 comments, and 12,369 comments, respectively.

\textbf{Mining reviewer experience.} 
Since reviewers have different ACO and RSO profiles at each point in time, we retrieved their authoring and reviewing histories with respect to each PR in the dataset.
For ACO of a reviewer $D$ for a PR $p$, we used PyDriller~\cite{PyDriller} to retrieve the number of previous commits that $D$ authored and the total number of commits in the corresponding repository.
For RSO of a reviewer $D$ for a PR $p$, we used the GitHub search API via their GraphQL implementation to retrieve the number of previous PRs in which the reviewer $D$ participated and the total number of PRs in the corresponding repository.
We capture all previous commits and PRs that were submitted before the submission of a PR $p$ since the inception of the repository.
% For ACO, we used PyDriller \cite{PyDriller} to retrieve commit histories from 826 repositories.
% We use the name and timestamp information from the commits to calculate the ACO for every single example in the dataset.
% For RSO, we utilised the GitHub search API via their GraphQL implementation to retrieve review histories.  
% Similarly, we used the username and timestamp from the reviews to calculate the RSO for every single example in the dataset.

\begin{table}[t]
\caption{Transformation of Training(Tr), Validation(Val) and Test(Te) sets.}
\label{tab:data_transformation}
\begin{tabular}{l
>{\columncolor[HTML]{EFEFEF}}l 
>{\columncolor[HTML]{C0C0C0}}l 
>{\columncolor[HTML]{9B9B9B}}l }
                            & \textbf{Tr}      & \textbf{Val}    & \textbf{Te}     \\ \hline
Original Size               & 150,406 & 13,103 & 13,104 \\ \hline
Deleted Reviews             & 618     & 24     & 41     \\
Bot Reviews                 & 1,207   & 41     & 55     \\
No Natural Language Comment & 7,322   & 632    & 639    \\ \hline
Final Size                  & 141,259 & 12,406 & 12,369 \\ \hline
\end{tabular}
\end{table}

\begin{table}[t]
\caption{Accounts in Training(Tr), Validation(Val) and \\Test(Te) sets.}
\label{tab:accounts}
\begin{tabular}{l
>{\columncolor[HTML]{EFEFEF}}l 
>{\columncolor[HTML]{C0C0C0}}l 
>{\columncolor[HTML]{9B9B9B}}l }
                        & \textbf{Tr} & \textbf{Val} & \textbf{Te} \\ \hline
Reviewer Accounts       & 10,583      & 2,148        & 2,125       \\ 
Identified Bot Accounts & 9           & 3            & 4           \\ \hline
\end{tabular}
\end{table}

\subsection{Experimental Setup}
% This section provides details to our experiment setup.

\textbf{Model Selection.}
We experiment with the state-of-the-art automated code review model, CodeReviewer~\cite{li2022automating}. 
This model is a 225M parameter transformer which was pre-trained from CodeT5's weights~\cite{wang2021codet5}.
% The model was pre-trained using \textcolor{red}{XXX} GitHub code reviews data.
The code review oriented pre-trained model was then fine-tuned into three standalone models for different review tasks.
% , i.e., quality estimation, review comment generation, and code refinement.
In this work, we focus on the review comment generation model.
Since we repurpose the code refinement dataset to the comment generation task, we replicate the original comment generation model using our newly prepared dataset for a fair comparison in our experiment.
Note that although the model is re-finetuned, we achieve a similar performance (BLEU-4 of 7.27) as in the original paper (BLEU-4 of 5.32)~\cite{li2022automating}.
% After being initialized with CodeT5's \cite{wang2021codet5} weights, CodeReviewer was continually trained on the tasks of \textit{Diff Tag Prediction}, \textit{Denoising Code Diff}, \textit{Denoising Review Comment} and \textit{Review Comment Generation}.
% The pre-training dataset consisted of GitHub code reviews that were collected together with the finetuning dataset mentioned above.
% Whilst CodeReviewer is capable of performing \textit{Code Diff Quality Estimation} and \textit{Code Refinement}, we focus on its \textit{Code Review Generation} capability for the purpose of our study.

\textbf{Experience-aware oversampling.}
We explore different types of experienced reviewers: a) Major Reviewer Major Authors only (\textbf{\textit{MRMA}}), b) all Major Reviewers (\textbf{\textit{MR}}) and c) all Major Authors (\textbf{\textit{MA}}). Table \ref{tab:exp_dist} shows a proportion of examples in each experience type.
We fine-tune the pre-trained CodeReviewer model by targeting one of the three types.
To do so, we upsample the subset of reviews associated with the target experience type by 400\% in the training data to achieve a 2:3 ratio for the smallest partition~\cite{przystupa-abdul-mageed-2019-neural}.
% , we experimented with oversampling different subsets of the data.
% These subsets are a) Major Reviewer Major Authors only (\textit{MRMA}), b) all Major Reviewers (\textit{MR}) and c) all Major Authors (\textit{MA}).
% Treating the scenario as a low resource neural machine translation problem \cite{currey-etal-2020-distilling, przystupa-abdul-mageed-2019-neural,tan-etal-2019-multilingual}, we upsample the target demographic three times and concatenate with the rest of the data for each approach.

\textbf{Settings.}
We fine-tune the original and the three oversampling models (i.e., MRMA, MR, MA) using the same hyperparameters as Li et al.~\cite{li2022automating}, i.e., a batch size of 72, learning rate of 0.0003, beam search width of 10, and trained the model for 30 epochs.
For hardware, we used a 32-core server with four NVIDIA A100-80G GPUs.

\subsection{Evaluation}
We used one quantitative measure and five human evaluation tasks:

\begin{itemize}
    \item \textbf{BLEU-4 (RQ1)}: Similar to~\citet{li2022automating}, we use BLEU-4~\cite{bleu} to assess the deviations in performance as a canonical benchmark.
    \item \textbf{Sematic Equivalence (RQ1)}: We manually examined whether the generated comments are semantically equivalent, i.e., same intention as the ground truth, regardless of the degree of textual overlap~\cite{tufano2022using, li2022automating}.
    \item \textbf{Applicability (RQ2)}: We manually determine that the generated comment is considered applicable if it raises a valid suggestion or concern in the context of the submitted PR, regardless of its semantic equivalence with the ground truth.
    % For each upsample strategy, we record the amount of generated reviews that are semantically equivalent to the ground truth (\textit{GT}).
    \item \textbf{Feedback type (RQ2)}: We categorise the comments into three distinct types: \textit{Suggestion} (i.e., proposing a solution), \textit{Concern} (i.e., raising issues), and \textit{Confused Question} (i.e., showing a lack of understanding or a need for clarification).
% For a comment to be regarded as a suggestion, it needs to propose a solution, comments that merely raise issues are concerns.
% Lastly, confused questions represent a lack of understanding and a need for clarification.
    \item \textbf{Presence of explanation (RQ2)}: We examine whether a comment includes its rationale or explanation. 
    \item \textbf{Comment Category (RQ3)}: We categorise a comment based on 18 categories developed by past work~\cite{samsung,characteristics_microsoft,mantyla}: \textit{Larger Defect}, \textit{Validation}, \textit{Logical}, \textit{Interface}, \textit{Solution Approach}, \textit{Question}, \textit{Design Discussion}, \textit{Resource}, \textit{Documentation}, \textit{Organization of Code}, \textit{Alternate Output}, \textit{Support}, \textit{Timing}, \textit{Naming Convention}, \textit{Praise}, \textit{Visual Representation}, \textit{False Positives}, and \textit{others}.

\end{itemize}

We measure BLEU-4 on the entire test set.
The manual evaluation tasks were conducted on 100 samples in the test set, which should allow us to generalise conclusions with a confidence level of 95\% and a confidence interval of 10\%. 
The first and fourth authors evaluated 25 samples separately, achieving Cohen's kappa ($\kappa$) between [0.28, 0.45] for semantic equivalence, [0.12, 0.35] for applicability, [0.52, 1] for feedback type, [0.46, 0.63] for explanation, and [0.17, 0.33] for comment category.
After resolving the discrepancies, the first author continued to classify the remaining 75 samples.
The classification of these 75 samples are then reviewed by the fourth author to ensure the consistency of manual evalution.

\begin{table}[t]
\caption{Distribution of Examples in the Training (Tr), Validation (Val), and Test (Te) sets.}
\label{tab:exp_dist}
\begin{tabular}{lcccccc}
\hline
\multicolumn{1}{|l|}{}             & \multicolumn{3}{l|}{\textbf{Major Reviewer}}                                                                                                    & \multicolumn{3}{l|}{\textbf{Minor Reviewer}}                                                                             \\ \hline
\multicolumn{1}{|l|}{\textbf{Major Author}} & \cellcolor[HTML]{EFEFEF}14\% & \cellcolor[HTML]{C0C0C0}{\color[HTML]{000000} 40\%} & \multicolumn{1}{l|}{\cellcolor[HTML]{9B9B9B}42\%} & \cellcolor[HTML]{EFEFEF}7\% & \cellcolor[HTML]{C0C0C0}18\% & \multicolumn{1}{l|}{\cellcolor[HTML]{9B9B9B}19\%} \\ \hline
\multicolumn{1}{|l|}{\textbf{Minor Author}} & \cellcolor[HTML]{EFEFEF}21\% & \cellcolor[HTML]{C0C0C0}18\%                        & \multicolumn{1}{l|}{\cellcolor[HTML]{9B9B9B}17\%} & \cellcolor[HTML]{EFEFEF}58\% & \cellcolor[HTML]{C0C0C0}24\% & \multicolumn{1}{l|}{\cellcolor[HTML]{9B9B9B}22\%} \\ \hline
& \cellcolor[HTML]{EFEFEF}\textbf{\textit{Tr}} & \cellcolor[HTML]{C0C0C0}\textbf{\textit{Val}} & \cellcolor[HTML]{9B9B9B}\textbf{\textit{Te}}& \cellcolor[HTML]{EFEFEF}\textbf{\textit{Tr}} & \cellcolor[HTML]{C0C0C0}\textbf{\textit{Val}} & \cellcolor[HTML]{9B9B9B}\textbf{\textit{Te}}\\

\end{tabular}
\end{table}

\section{Results}

\begin{figure*}[ht] 
    \centering
    \includegraphics[width=\textwidth, trim=0 25 0 0, clip]{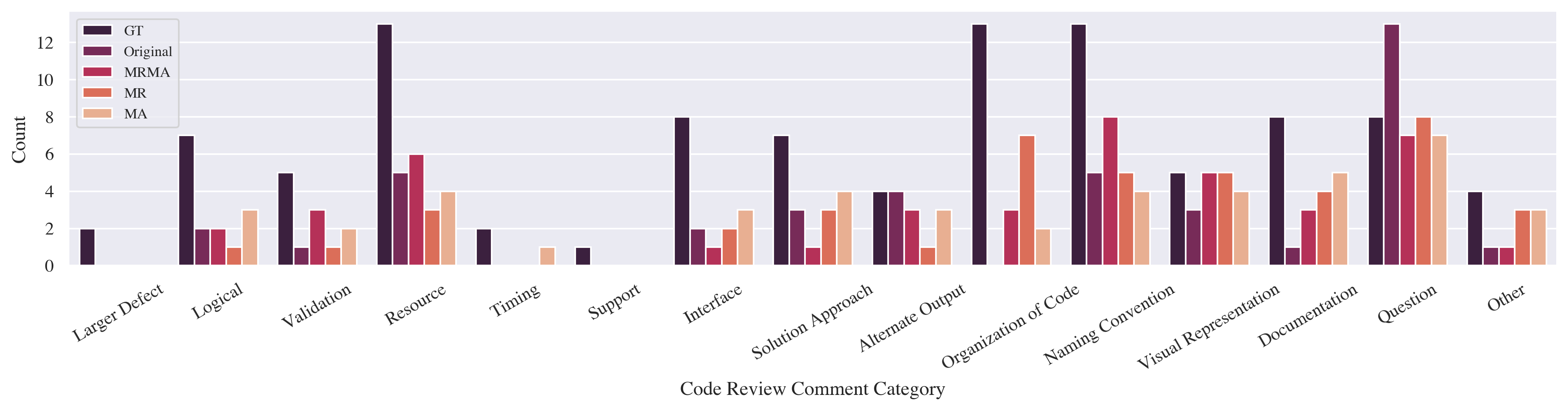} 
    \caption{Comment Categories of Applicable Code Review Comments}
    \label{fig:comment_cat}
\end{figure*}

\textbf{RQ1 - \rqone}
Table \ref{tab:rq1_result} shows the average BLEU-4 and the number of comments that are semantically equivalent to the actual comments by any reviewers (All), by major reviewer major authors ($\lozenge$$\blacklozenge$), by major reviewers ($\lozenge$), and by major authors ($\blacklozenge$) in the test set.
The results show that although the BLEU-4 scores of our oversampling models are slightly lower than those of the original model, all of our oversampling models achieve a higher number of comments that are semantically equivalent to the ground truth compared to the original model.
The lower BLEU is due to harsher penalties to any differences in short sentences~\cite{song2013bleu,fomicheva2019taking}. Prior work also report that BLEU is known to poorly correlate with human judgement~\cite{callison2006re,belz2006comparing,tan2015awkward,smith2016climbing,kann2018sentence} and ignore the semantic quality~\cite{babych2008sensitivity}.
% and is unsuitable for short texts \cite{song2013bleu,fomicheva2019taking}.
Specifically, the \textit{MA} model achieves the highest number of semantically equivalent comments.
The findings highlight a better alignment with experienced reviewers' perspectives in our models.

\begin{table}[t]
\setlength\heavyrulewidth{0.25ex}
\caption{BLEU-4 (B4) on the Test Set \& Semantic Equivalence (SE) on the 100 Samples}
\label{tab:rq1_result}
\resizebox{\columnwidth}{!}{
\begin{tabular}{
  @{}
  S[table-format=1.2]
  S[table-format=1.2]
  S[table-format=1.2]
  S[table-format=1.2]
  S[table-format=1.2]
  S[table-format=-1.2]
  S[table-format=1.2]
  S[table-format=1.2]
  S[table-format=1.2]
  S[table-format=1.2]
  @{}
}

\toprule
 & & & \multicolumn{6}{c}{Oversampling models} \\
\cmidrule(l){4-9}
 & \multicolumn{2}{c}{Original} & \multicolumn{2}{c}{MRMA} & \multicolumn{2}{c}{MR} & \multicolumn{2}{c}{MA} \\
\midrule
& {\textit{B4}} & {\textit{SE}} & {\textit{B4}} & {\textit{SE}} & {\textit{B4}} & {\textit{SE}} & {\textit{B4}} & {\textit{SE}}\\

All &  {7.27} & {15/100} & {7.12} & {18/100} & {7.1} & {16/100} & {7.11} & {21/100} \\
$\lozenge$$\blacklozenge$  & {6.99} & {6/32} & {6.78} & {6/32} & {6.94} & {7/32} & {6.85} & {11/32} \\
$\lozenge$ &  {7.09} & {7/52} & {6.93} & {10/52} & {6.98} & {9/52} & {6.89} &{14/52} \\
$\blacklozenge$ &  {7.12} & {9/48} & {6.9} & {10/48} & {6.99} & {11/48}  & {7} & {15/48} \\ \bottomrule
\multicolumn{9}{l}{\footnotesize $\lozenge$$\blacklozenge$ comments by Major Reviewer Major Authors in test set} \\
\multicolumn{9}{l}{\footnotesize$\lozenge$ comments by Major Reviewers \& $\blacklozenge$ comments by Major Authors in test set }
\end{tabular}
}
\vspace{-3mm}
\end{table}
% upsampling strategies have degraded results for all partitions of the test set compared to the original model, however the differences are negligible. 
% The largest gap in terms of the entire test set is between \textit{MA} and the original approach (-0.16).

% \textbf{Semantic Equivalence.}
% Based on our sample, we find that upsampling strategies have higher amounts of comments that are semantically equivalent to the ground truth compared to the original approach for all partitions.
% The largest gap in terms of the entire test set is between \textit{MA} and the original approach (+9).

\textbf{RQ2 - \rqtwo{}}
Table \ref{tab:rq2_result} shows that all of our oversampling models have produced more applicable comments than the original model.
Based on the applicable comments, we further evaluate the level of information.
We find that our oversampling models provide more comments with suggestions and less concerns than the original model.
On the contrary, only 55\% of the original approach's applicable comments were suggestions.
The original model has a higher tendency to generate confused questions (6/40) than our oversampling models, which generate 1--4 confused questions.
We also find that the original model seldom explains (4/40), whilst our oversampling models justified themselves more.
% Nevertheless, all generated approaches explained themselves far less than found in the ground truth (68/100).
Our results suggest that our oversampling models exhibit a notable increase in solution-oriented suggestions with explanation, whilst showing less confusion, which is a common review anti-pattern~\cite{antipattern}.

% \textbf{Applicability.}
% We count the amount of applicable comments generated by each upsample strategy.
% As shown in Table 3, all upsampling strategies have produced more applicable reviews than the original approach. 
% The largest gap in terms of the entire test set is between \textit{MA} and the original approach (+6).
% \textbf{Feedback type.} 
% Table 3 shows that the original approach has the highest tendency of expressing confused questions (6/38), as opposed to (4/42), (1/42) and (2/44) by \textit{MRMA}, \textit{MR} and \textit{MA}, respectively.
% Upsampled strategies provided more suggestions, with (33/42) \textit{MR} reviews providing a solution. 
% Contrarily, only half of the original approach's applicable comments were suggestions.
% Compared to all generated reviews, the ground truth still provided the highest rate of suggestions (77/100), whilst exhibiting the lowest rate of confusion (3/100).

% \textbf{Presence of explanation.} 
% If a review provides a rationale for their suggestion, it is considered to have provided an explanation.
% We measure the frequency in which the ground truth and each of the generated reviews attempt to explain themselves.
% We find that the original approach seldom provides an explanation (4/38), whilst \textit{MRMA} (15/42), \textit{MR} (16/42) and \textit{MA} (9/44) justified themselves more.

% Nevertheless, all generated approaches explained themselves far less than found in the ground truth (68/100).

\textbf{RQ3 - \rqthree{}}
Figure \ref{fig:comment_cat} shows that our models can generate new comments on functional issues e.g.,\textit{Logical}, \textit{Validation}, \textit{Resource}, despite that they are rarely raised \cite{microsoftbugs}.
In terms of maintenance-related issues, the \textit{MR} model excelled in this domain, while the \textit{MRMA} model displayed a higher propensity to suggest \textit{Naming Convention} related fixes.
On the other hand, the original model could not provide any \textit{Organization of Code} related feedback.
Moreover, the original model tends to ask questions more often. 
The results also highlight that our oversampling techniques can elicit more critical types of comments.

% In terms of functional issues, reviews were seldom related to \textit{Larger Defect}, \textit{Timing} and \textit{Support}. 
% Whilst all generated approaches found a similar amount of \textit{Logical}, \textit{Resource} and \textit{Interface} related issues, the original approach failed to find any \textit{Validation} issues.

% In terms of maintenance related issues, The original approach could not provide any \textit{Organization of Code} related feedback, whilst \textit{MR} excelled in this domain.
% Additionally, \textit{MRMA} displayed a higher propensity to suggest \textit{Naming Convention} related fixes.

% As seen in Figure 1, the original approach tends to ask questions more often. 
% These are partially concerns with the code under review and partially displays of an inability to comprehend the change.

% We discard the \textit{false positives} class as these are simply the results that were not applicable.
% \textit{Praise} and \textit{Design Discussion} did not appear in our sample, most likely due to the fact that the dataset only includes changes triggered by a single review comment.

\begin{table}[t]
\setlength\heavyrulewidth{0.25ex}
\caption{Human Evaluation of Applicability, Feedback Type, and Presence of Explanation on the 100 Samples}
\label{tab:rq2_result}
\begin{tabular}{
  @{}
  l
  S[table-format=1.2]
  S[table-format=1.2]
  S[table-format=1.2]
  S[table-format=1.2]
  S[table-format=-1.2]
  S[table-format=1.2]
  S[table-format=1.2]
  S[table-format=1.2]
  S[table-format=1.2]
  @{}
}
\toprule
 & & & \multicolumn{3}{c}{Oversampling models} \\
\cmidrule(l){4-6}
 & {GT} & {Original} & {MRMA} & {MR} & {MA} \\
\midrule
Applicability & {100} &  {40} & {43} & {43} & {45} \\
\midrule
Suggestion & {77} &  {22} & {32} & {34} & {32} \\
Concern & {20} &  {12} & {7} & {8} & {11} \\
Confused Question & {3} &  {6} & {4} & {1} & {2} \\
\midrule
Explanation & {68} &  {4} & {15} & {16} & {9} \\
\bottomrule
\end{tabular}
\end{table}

\section{Threats to Validity}
% \textbf{Internal validity.} 
We use the same pre-trained model, hyper-parameters, and training setup as CodeReviewer~\cite{li2022automating} so that the only varying factor is the fine-tuning dataset and our oversampling strategies.
To ensure the validity of the results, we replicate the original model and confirm that the model performance is similar to the original paper's results (5.35 vs 5.32) based on BLEU-4 on their test set.

Experience metrics may be underestimated when reviews or commits are deleted, users are unsearchable or use multiple accounts.
We treat these false negatives as noise, which should only under represent the potential impact of our technique.

Selected upsampling ratios are arbitrary and do not reflect the best upsampling performance that can be achieved in this dataset.
We leave tuning for the optimal ratio to future work.

The outcomes of the manual evaluations are prone to subjective perspectives of human evaluators.
To mitigate this, two annotators independently evaluate the sample and discuss to (1) resolve all disagreements and (2) apply the shared understanding to the rest of the annotation.
We include annotation results in the materials for transparency.
The data is subject to the confines of GitHub. 
As such, the targeted reviewer demographic does not reflect their counterparts in other software development environments.

% \textbf{Construct validity.} 
% The quality of the generated review comments cannot be comprehensively reflected with only BLEU-4 \cite{bleu} and the ground truth examples, hence we measure five different human evaluation metrics to better capture these results.

% \textcolor{red}{talk about other bot detection can be used but ...}
% As advanced methods \cite{botpattern,bothunter,golzadeh2021identifying, golzadeh2021ground,EnsBoD} for bot detection rely on features that are difficult to retrieve at this scale and cannot guarantee perfect precision

\section{Conclusion}
This study explores the ability of the automated code review model to generate higher-quality reviews by oversampling experienced reviewers' examples within the training set. 
Our results show that experience-aware oversampling allowed the model to generate more semantically correct comments and convey better information by providing more suggestions and explanations.
Additionally, the model was able to generate more comments related to functional issues.
As the underlying dataset is fundamentally unchanged, we demonstrate the existence of untapped knowledge within the experienced reviewer partitions of the training data.
% proving that the model has the capabilities to deliver more meaningful reviews that developers are looking for.

In future work, we intend to better understand the behavioural differences caused by experience-aware oversampling by investigating changes in attention weights.
We plan to experiment with different training methods to learn more effectively from experienced reviewers. Additionally, we will also compare with oversampling of novice reviewers to investigate if contrasting effects arise.

% As minor reviewer minor authors are the overwhelming majority of the training set, it is crucial to ensure that experienced reviewers' perspectives are not underrepresented.

\textbf{Data Availability.}
All the materials produced from this study are available on Zenodo\footnote{\url{https://zenodo.org/records/10572047}}.

\section*{Acknowledgment}
This research was supported by The University of Melbourne’s Research Computing Services and the Petascale Campus Initiative.
Patanamon Thongtanunam was supported by the Australian Research Council's Discovery Early Career Researcher Award (DECRA) funding scheme (DE210101091).

\balance
\bibliographystyle{ACM-Reference-Format}
\bibliography{references}

\end{document}